\documentclass[10pt,journal,final]{IEEEtran}
\usepackage{cite}
\usepackage{scalerel}
\usepackage{tikz}
\usetikzlibrary{svg.path}
\usepackage{epstopdf}
\usepackage{multirow}

\usepackage{amsmath,amssymb,amsfonts}
\usepackage{algorithmic}
\usepackage{graphicx}
\usepackage{subfigure}
\usepackage{booktabs}  
\usepackage{lipsum} 
\usepackage{textcomp}
\usepackage{wrapfig}
\usepackage{multirow}
\usepackage{color}
\usepackage{hhline}
\definecolor{RedColor}{rgb}{1,0,0}
\definecolor{BlueColor}{rgb}{0,0,1}
\definecolor{GreenColor}{rgb}{0,1,0}
\usepackage[colorlinks,            linkcolor=red,            anchorcolor=blue,            citecolor=green            ]{hyperref}
\definecolor{orcidlogocol}{HTML}{A6CE39}
\tikzset{
	orcidlogo/.pic={
		\fill[orcidlogocol] svg{M256,128c0,70.7-57.3,128-128,128C57.3,256,0,198.7,0,128C0,57.3,57.3,0,128,0C198.7,0,256,57.3,256,128z};
		\fill[white] svg{M86.3,186.2H70.9V79.1h15.4v48.4V186.2z}
		svg{M108.9,79.1h41.6c39.6,0,57,28.3,57,53.6c0,27.5-21.5,53.6-56.8,53.6h-41.8V79.1z M124.3,172.4h24.5c34.9,0,42.9-26.5,42.9-39.7c0-21.5-13.7-39.7-43.7-39.7h-23.7V172.4z}
		svg{M88.7,56.8c0,5.5-4.5,10.1-10.1,10.1c-5.6,0-10.1-4.6-10.1-10.1c0-5.6,4.5-10.1,10.1-10.1C84.2,46.7,88.7,51.3,88.7,56.8z};
	}
}

\newcommand\orcidicon[1]{\href{https://orcid.org/#1}{\mbox{\scalerel*{
				\begin{tikzpicture}[yscale=-1,transform shape]
					\pic{orcidlogo};
				\end{tikzpicture}
			}{|}}}}

\usepackage{hyperref} 

\usepackage{graphicx}
\usepackage{float} 
\usepackage{subfigure}
\usepackage{color}
\usepackage{amsmath}

\begin{document}
%
\title{Robust PCA Unrolling Network for Super-resolution Vessel Extraction in X-ray Coronary Angiography}
%
%
%

\author{Binjie Qin* \orcidicon{0000-0001-7445-1582}, Haohao Mao \orcidicon{0000-0001-9730-6062}, Yiming Liu \orcidicon{0000-0002-7016-6415}, Jun Zhao, Yisong Lv \orcidicon{0000-0001-7091-2062}, Yueqi Zhu, Song Ding, Xu Chen \orcidicon{0000-0002-4870-2381}
\thanks{The manuscript was received on April 14, 2021, and accepted on October 27, 2018. This work was partially supported by the Science and Technology Commission of Shanghai Municipality (19dz1200500, 19411951507), the National Natural Science Foundation of China (61271320,82070477), Shanghai ShenKang Hospital Development Center (SHDC12019X12), and Interdisciplinary Program of Shanghai Jiao Tong University (ZH2018ZDA19,YG2021QN122,YG2021QN99).(Corresponding author: Binjie Qin.)}
\thanks{Binjie Qin, Haohao Mao, Yiming Liu, and Jun Zhao are with School of Biomedical Engineering, Shanghai Jiao Tong University, Shanghai 200240, China. E-mail: bjqin@sjtu.edu.cn }
\thanks{Yisong Lv is with School of Continuing Education, Shanghai Jiao Tong University, Shanghai 200240, China.}
\thanks{Yueqi Zhu is with Department of Radiology, Shanghai Jiao Tong University Affiliated Sixth People’s Hospital, Shanghai Jiao Tong University, 600 Yi Shan Road, Shanghai 200233, China.}
\thanks{Song Ding is with Department of Cardiology, Ren Ji Hospital, School of Medicine, Shanghai Jiao Tong University, Shanghai 200127, China.}
\thanks{Xu Chen is with Center for Advanced Neuroimaging, University of California, Riverside, 900 University Avenue, Riverside, CA 92521, USA.}
}
%
%

\markboth{Revision Submitted to IEEE Transactions on Medical Imaging, 14-Mar-2022}%
{Shell \MakeLowercase{\textit{et al.}}: Bare Demo of IEEEtran.cls for IEEE Journals}
%



\maketitle

\begin{abstract}
Although robust PCA has been increasingly adopted to extract vessels from X-ray coronary angiography (XCA) images, challenging problems such as inefficient vessel-sparsity modelling, noisy and dynamic background artefacts, and high computational cost still remain unsolved. Therefore, we propose a novel robust PCA unrolling network with sparse feature selection for super-resolution XCA vessel imaging. Being embedded within a patch-wise spatiotemporal super-resolution framework that is built upon a pooling layer and a convolutional long short-term memory network, the proposed network can not only gradually prune complex vessel-like artefacts and noisy backgrounds in XCA during network training but also iteratively learn and select the high-level spatiotemporal semantic information of moving contrast agents flowing in the XCA-imaged vessels. The experimental results show that the proposed method significantly outperforms state-of-the-art methods, especially in the imaging of the vessel network and its distal vessels, by restoring the intensity and geometry profiles of heterogeneous vessels against complex and dynamic backgrounds.

\end{abstract}

\begin{IEEEkeywords}
Algorithm unrolling, RPCA unrolling network, X-ray coronary angiography, Vessel extraction, Sparse feature selection, Super-resolution.
\end{IEEEkeywords}

%
\IEEEpeerreviewmaketitle

\section{Introduction}
%
%
%
%
\IEEEPARstart{C}{ardiovascular} diseases (CVDs) threaten human health worldwide\cite{sechtem2020coronary}. Percutaneous coronary intervention (PCI) is very important for the diagnosis and treatment of CVDs, during which X-ray coronary angiography (XCA) is a primary technique for imaging morphological and functional information about blood vessels. Due to X-ray beams being attenuated by varying amounts when they pass through tissues with different densities along the projection path of XCA imaging, the XCA sequence displays heterogeneous blood vessels that overlap with various anatomical structures (such as bones, lungs and diaphragms), mixed Poisson-Gaussian noise\cite{Irrera2016flexible,zhao2019texture}, and respiratory and cardiac motions. It is very difficult for surgeons to clearly identify blood vessels, let alone extract vessels for the quantitative analysis of a vessel's structure and function. Vessel extraction algorithms \cite{moccia2018blood,jia2021learning} are usually built upon tube-like feature representation, which is very sensitive to noisy and dynamic background artefacts. Recently, XCA vessel extraction\cite{2017Extracting,ma2017automatic,jin2018low,qin2019accurate,fang2019topology,song2020spatio,xia2020vessel} was regarded as the separation of foreground vessels and background structures within sparse and low-rank modelling via robust principal component analysis (RPCA)\cite{bouwmans2018applications} to achieve state-of-the-art performance. However, the extracted results in these studies still have some noisy artefacts.

Moreover, RPCA-based vessel extraction consumes a large amount of storage and time. Therefore, a deep neural network called a convolutional robust PCA (CORONA)\cite{solomon2020deep} unfolds the RPCA algorithm for ultrasonic vascular imaging. Algorithm unrolling\cite{monga2021algorithm} or unfolding was first introduced in \cite{gregor2010learning}, the result of which being that the time efficiency of the unfolded deep network is greatly improved compared with that of the original iterative algorithm. However, the problems caused by the dynamic background and complex noise patterns still remain in the CORONA solution when it is used to extract vessels from XCA images. Furthermore, the mixed Gaussian-Poisson noise in XCA is complex and heterogeneous in different XCA imaging machines and locally affects the extraction of distal vessels with low contrast and low SNR. RPCA-based methods and CORONA globally implement foreground/background decomposition and cannot effectively cope with the local interference caused by mixed noise and heterogeneous artefacts. 

To solve these problems, we propose a patch-wise spatiotemporal super-resolution (SR) module to refine the vessel features outputted by the deep unfolded RPCA layer. Specifically, a feature pooling layer inputs the original data to the unfolded RPCA layer and applies patch-wise sparse feature selection in the SR module to eliminate redundant vessel-like artefacts while retaining the useful vessel features \cite{singh2020eds}. Different from current deep-learning-based SR neural networks\cite{wang2021deep} that automatically extract features for a non-linear low-resolution to high-resolution mapping and cannot select local/non-local sparse features from a single image or video, the proposed RPCA unrolling network, called RPCA-UNet, implements a patch-wise spatiotemporal SR module with sparse feature selection that is based on a residual module and a convolutional long short-term memory (CLSTM) network \cite{shi2015convolutional}. RPCA-UNet can effectively enhance patch-wise vessel features by extracting not only the heterogeneous grey level information but also the geometrical structures of XCA vessels in a spatiotemporally consistent way. Specifically, the residual module is first applied to extract deep features through multiple convolutional layers and transfer these features along with the original features via the residual operation to the subsequent CLSTM network. By saving complementary features of pervious frame in an XCA sequence, the CLSTM network integrates the features of the current frame into the complementary features of previous frames. This feature aggregation establishes a spatiotemporal evolution for accurately extracting both the image grey values and geometrical features of XCA vessels. The main contribution of this work is threefold:
\begin{enumerate}
	\item A novel RPCA unrolling (or unfolded RPCA) network with a patch-wise SR module is proposed to iteratively extract XCA vessels with a certain time and space efficiency. The unrolling network in each iteration/layer has a pooling layer as the preprocessing layer and a patch-wise SR module as the postprocessing layer that consists of a residual module and a CLSTM network. Our proposed RPCA-UNet can not only achieve uninformative feature pruning and Gaussian-Poisson denoising but also selectively learn sparse vessel features from complex and dynamic backgrounds. To the best of our knowledge, this is the first use a RPCA unrolling network with sparse feature selection to extract vessels from XCA images. Experiments show that the proposed method significantly outperforms state-of-the-art methods in both vessel extraction and vessel segmentation.	
	\item We apply CLSTM network to the proposed SR module that can not only learn sparse features selectively from the current frame of XCA sequence but also preserve the high-level spatiotemporal semantic detail of moving contrast agents in the whole XCA sequence. CLSTM network in a patch-wise SR network is proven to boost the performance of vessel extraction by significantly improving the distal vessel detection accuracy and spatiotemporal consistency in the XCA sequence.
	\item The proposed RPCA-UNet is implemented as a weakly supervised learning method such that grey value vessel labelling is automatically generated by our vessel extraction method, called VRBC (vessel region background completion)\cite{qin2019accurate}, and our training data and testing data comprise heterogeneous XCA images that are collected from different machines. This weakly supervised learning in a heterogeneous environment overcomes the need of expensive and time-consuming manual annotation and improves a generalization ability of the proposed network.
\end{enumerate}

\section{Related Works}
\subsection{XCA Vessel Extraction}
Compared with other imaging modalities reviewed in recent survey studies\cite{moccia2018blood,jia2021learning,mookiah2021review}, such as computed tomography angiography, magnetic resonance angiography and retinal fundus images, few studies on extracting vessels from XCA images have been conducted. XCA vessel extraction methods\cite{qin2022extracting} can be divided into the following four categories that corporately transform XCA images into segmentation results: vessel enhancement, deformable model, vessel tracking, and machine learning. Vessel enhancement approaches \cite{cervantes2018coronary,fazlali2018vessel,kerkeni2016coronary,wan2018automated} aggregate compact image patches in local/non-local filtering to enhance the tube-like vessel features and increase the contrast between the foreground vessels and background structures. For example, Hessian-based multiscale local or non-local\cite{qian2021vascular} filtering yielding geometrical features \cite{frangi1998multiscale} regarding both vesselness and direction information was incorporated into an iterative region growing \cite{kerkeni2016coronary}, a statistical region merging \cite{wan2018automated}, and multiscale superpixels\cite{fazlali2018vessel} to group enhanced pixels into correct clusters of arteries and background. However, Hessian-matrix-based segmentation is largely dependent on the optimal scale selection of the major vessel radius and highly sensitive to mixed Gaussian-Poisson noise in the spatial domain. In the frequency domain, single-scale Gabor filters with optimized parameters\cite{cervantes2018coronary} and multiscale Gabor filters with optimized response thresholding \cite{cruz2016automatic} are developed for XCA vessel segmentation, but the thresholding techniques in the Gabor and wavelet domains cannot easily distinguish the vascular structures from many vessel-like artefacts. Inspired by phase congruency, which has stability in the presence of noise and is invariant to changes in contrast, some detail-preserving image enhancement methods have exploited phase-congruency-based feature indicator called phase symmetry or phase asymmetry\cite{zhao2018automatic,mei2019phase} to detect edge- and ridge-like features such as 2D/3D vessels\cite{zhao2018automatic,reisenhofer2019edge}. Nevertheless, vessel enhancement methods can simultaneously enhance the vessel-like background structures in XCA images.

Another common method is using deformable models to segment vessels, which have parametric and geometric deformable models. Parametric models such as active contour model \cite{kass1988snakes} directly represent the target curves or surfaces during deformation, resulting in an efficient and lower computational cost segmentation, but are not suitable for XCA vessels with complex topologies. In contrast, geometric deformable models are implemented as an energy minimization within a level set framework\cite{osher1988fronts,zou2021survey}, which can be adapted to changing vessel topologies and effectively extract thin vessels and complex vessel branches. Currently, integrating not only the edge and centerline information \cite{lv2019vessel} but also the region\cite{sun2012local} and shape prior constraints \cite{ge2019two} into the optimization model can lead to more precise XCA vessel segmentation. However, deformable models have strong dependence on initialization, high sensitivity to irregular vessel shapes with inhomogeneous intensity and low contrast, and high computational cost.

Additionally, vessel-tracking methods also attract much attention. Vessel-tracking methods usually place initial seed points and drive the growth process with specific constraints to segment the vessel area. They are generally divided into two categories by different tracking constraints: model-based \cite{fang2020greedy} and minimal path\cite{wink2004multiscale,chen2016curve,yang2020vessel} methods. Model-based techniques track vessels by searching for and matching a predefined vessel model with different shapes and thicknesses, but their matching performance decreases sharply on images with high noise and inhomogeneous intensity as well as many non-vascular structures. Minimal path methods \cite{wink2004multiscale,chen2016curve} can efficiently extract the XCA vessel centreline by finding the path with minimal accumulated cost between two given endpoints via centreline evolution over a filter-response-derived vectorial multiscale feature image \cite{wink2004multiscale} or via the backtracking operation\cite{chen2016curve}. The work in\cite{yang2020vessel} has extracted the complete vessel lumen within the framework of backtracked minimal path propagation. Due to a lack of global context feature selection, the centreline extraction method has difficulty in avoiding under-segmentation in some clear gaps between vessel structures with low-contrast-intensity inhomogeneities or in some weak vessel structures such as distal vessels.

Machine learning methods such as RPCA-\cite{2017Extracting,ma2017automatic,jin2018low,qin2019accurate,fang2019topology,song2020spatio,xia2020vessel} and graph-based\cite{liu2016robust,fang2020greedy} methods treat segmentation as an optimized classification to distinguish between foreground and background pixels\cite{jia2021learning}. However, the globally optimal solutions have several unsolved problems such as discriminative feature representation, spatiotemporal regularization, and mixed Gaussian-Poisson noise removal. With the ability to perform featurization and classification of big data, deep-learning-based methods, especially the convolutional neural network (CNN) combined with image enhancement\cite{nasr2018segmentation} for preprocessing and graph-based vessel connection\cite{shin2019deep} for post-processing as well as pyramid pooling and the convolving of multiscale features with small sample transfer learning\cite{zhu2021coronary}, have proven effective in XCA segmentation but still have several unsolved problems related to spatiotemporal and semantic context modelling. Using an encoder-decoder architecture equipped with skipping connections, U-Net\cite{ronneberger2015u} and fully convolutional networks (FCNs) combine high-level semantic information with low-level appearance details to efficiently achieve end-to-end semantic segmentation of entire vessel trees\cite{fan2018multichannel,hao2020sequential,wan2021automatic,zhu2021coronary,samuel2021vssc}. For example, SVS-net\cite{hao2020sequential} embedding channel attention mechanism for suppressing noisy backgrounds and the spatiotemporal FCN\cite{wan2021automatic} integrating interframe information with influence layers are proposed to extract multiscale features for segmenting entire vessels from XCA sequence. However, deep networks of a certain depth for accumulating multiscale feature have limitations in extracting more features and handling details such as distal vessels. 

Unfortunately, most deep learning techniques have limitations in not only efficiently extracting more spatiotemporal features in a sequential way but also discriminatively selecting sparse vessel features from vessel-like and signal-dependent noisy backgrounds. To the best of our knowledge, none of the current vessel extraction methods can fully restore the intensity and geometry profiles of entire heterogeneous XCA vessels, except VRBC method\cite{qin2019accurate}. 

\subsection{Unrolling Neural Network}
The unrolling neural network was first proposed by Gregor and LeCun \cite{gregor2010learning} to approximate the iterative soft-threshold algorithm (ISTA) for sparse coding. The unfolded network called the learned ISTA (LISTA) achieves great performance, being nearly 20 times faster than the traditional accelerated ISTA. The success of the LISTA shows the significant computational benefits of the deep unfolding algorithm. Moreover, by considering each iteration of an iterative algorithm as a layer of an unrolling network and then concatenating a few of these layers, one needs only a few iterations of training to achieve a dramatic improvement in convergence. 

Recently, algorithm unrolling has attracted significant attention in signal and image processing\cite{monga2021algorithm}, where the collection of sufficient data is expensive and difficult to achieve and the performance of conventional networks is limited. Solomon \textit{et al.} \cite{solomon2020deep} proposed CORONA to separate the blood vessels and background tissues from an ultrasound signal. CORONA was trained on simulated data, and then the resulting network was trained on in vivo data. This hybrid policy can not only improve the network performance but also achieve a fully automated network, in which all the regularization parameters are also learned. Moreover, by exploiting spatial invariance and facilitating the training process, one can reduce the number of learnable parameters dramatically through the use of convolutional layers. 

Algorithm unrolling shows great potential in solving inverse problems in biomedical imaging. Xiang \textit{et al.} \cite{xiang2021fista} unfolded the fast ISTA (FISTA) framework into FISTA-Net, which achieves great performance in different imaging tasks. Algorithm unrolling has been expanded to the graph domain and designed as an interpretable architecture from a signal processing perspective\cite{chen2021graph}. The graph unrolling networks\cite{chen2021graph} are trained through unsupervised learning, where the input noisy measurements are used to supervise the neural network training. The network output does not overfit the noisy input in most cases, indicating that the unrolling network can carry out implicit graph regularization and thus avoid overfitting.

\subsection{Feature Selection}
Feature selection \cite{cai2018feature} can improve learning accuracy while defying the curse of dimensionality of high-dimensional data in an efficient way. For example, PCI needs an accurate real-time navigation system to visualize and navigate inside the vessel network, which presents structure-functional information about the cardiac perfusion; thus, an XCA image could be categorized into vessel and non-vessel features in deep-learning-based image analysis. Therefore, feature selection\cite{cai2018feature} can be used to find the most appropriate lightweight feature subset that preserves relevant vessel information while discarding the non-vessel and artefact features. Generally, feature selection techniques can be classified into four main categories: filter, wrapper, embedded, and hybrid methods.

Filter methods evaluate feature relevance in discriminating different classes according to predefined criteria without using any learning algorithm. The criteria include information theoretic criteria such as mutual information \cite{battiti1994using} and multivariate joint entropy\cite{yu2020multivariate}. Filter methods are fast, but their selected subset is usually not an optimal feature subset from the perspective of classification performance. Recently, feature interactions among multiple variables \cite{yu2020multivariate} and views\cite{zhang2019feature} in multisource heterogeneous data environments were studied to increase the classification accuracy. 

Wrapper methods select features to achieve the best performance of a specific learning algorithm\cite{tarkhaneh2021novel,jiang2021wrapper}. Although various algorithms are used to accelerate the combinatorial optimization for maximizing the relevancy to the target class and minimizing the redundance of selected features, the computational cost of wrapper methods is still too high, especially when the number of selected features greatly increases. Therefore, hybrid approaches\cite{got2021hybrid} that use filter methods to remove the irrelevant features and then select important features from a candidate subset by wrapper models are developed to achieve the best possible performance by a particular learning algorithm with time complexity similar to that of the filter methods.

Because feature selection is embedded in the training of a learning machine, embedded methods are better than other methods in jointly achieving high classification performance and computational efficiency. Typically, the popular sparse learning models implement embedded feature selection \cite{gui2016feature} by minimizing an empirical error penalized by a regularization term such as the $l_{r,p}$-norm regularizer. For example, the traditional RPCA-based and CORONA methods usually apply the $l_{1,2}$-norm to select sparse features and eliminate redundant features. However, these approaches select solely individual sparse features and ignore the possible interaction between different features. Therefore, group sparse feature selection was recently developed\cite{li2020survey} to model the strengths of interactions between different features with graph-based edge weights and to partition the graph into different groups in terms of their strengths. However, how to combine sparse feature selection with neural networks\cite{zhao2015heterogeneous,farokhmanesh2021deep} is still a poorly understood and unexplored research area. While deep neural networks can automatically extract features appropriate for the target task and use an attention mechanism\cite{Dong2021Attention,hao2020sequential} to weigh the different features to increase the classification performance, they usually cannot check important input signals and select sparse features based on some predefined criteria, which leads to a lack of interpretability. To solve this problem, an attempt at feature selection is conducted in the proposed RPCA-UNet to gain prediction accuracy and computational efficiency with existing XCA data. Feature selection for pruning neural network and reinforcement learning is beyond the scope of our paper, we refer the interested reader to the related works\cite{Hoefler2021sparsity,Liu2022Automated}. 

\section{Method}
The overall architecture of each iteration/layer from RPCA-UNet is shown in \figurename{ 1} for decomposing a given XCA data $D$ into the sum of a vessel ($S$) and a background ($L$) component. It is difficult for data-driven neural networks to build models for removing the underlying mixed noises and artefacts. We then focus on sparse feature selection, which plays an important role in RPCA-UNet. Specifically, RPCA-UNet in each layer has a feature selection module that combines a pooling layer as the subsampling-based preprocessing layer with a patch-wise SR module as the post-processing layer, which consists of a convolutional layer, a residual module and a CLSTM network.
\begin{figure}[ht]
	\centering 
	\includegraphics[width=0.495\textwidth]{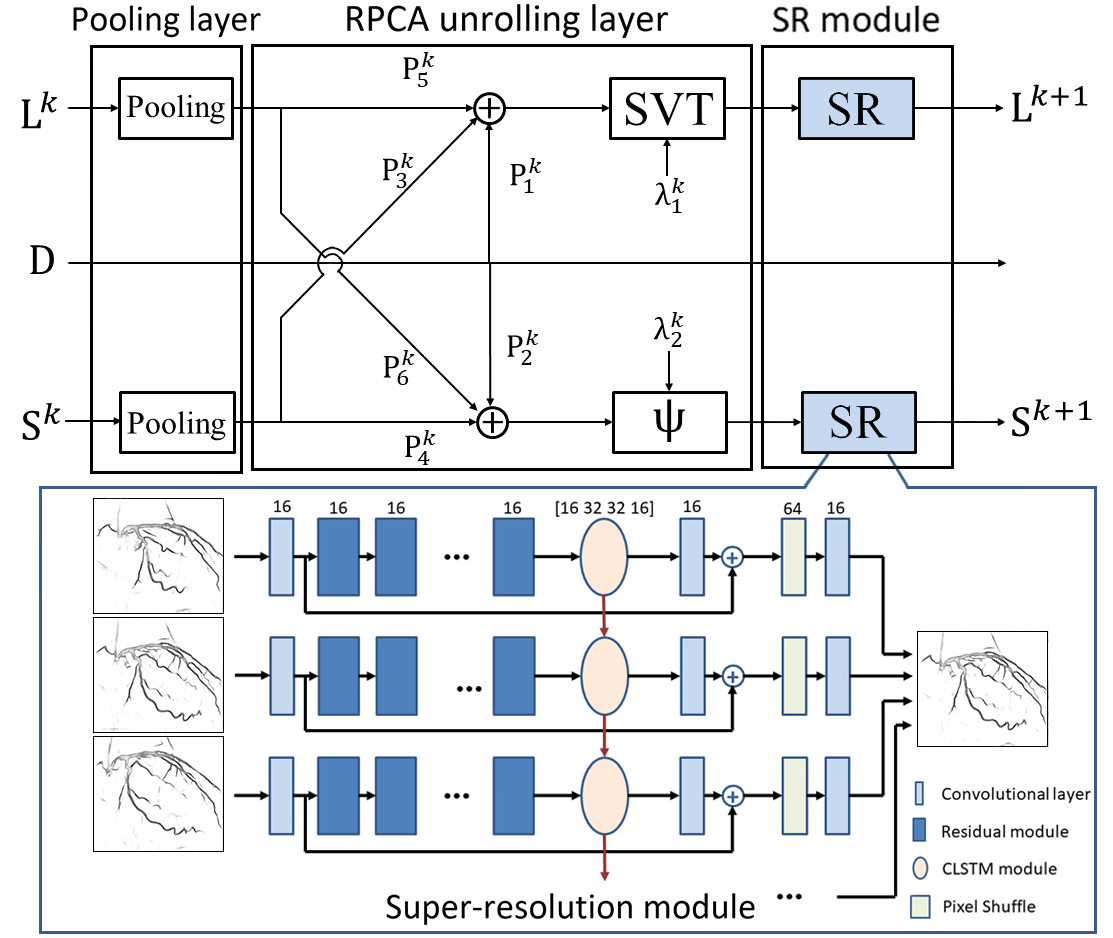}
	\caption{The architecture of a single iteration/layer of RPCA-UNet for decomposing XCA data $D$ into vessel ($S$) and background ($L$) components, which consists of a pooling layer, an RPCA unrolling layer, and an SR module. The SR module is mainly built upon the convolutional layer, residual module and CLSTM network. } 
	\label{2} 
\end{figure}

\subsection{RPCA Modelling}
The contrast agents move quickly in vessels and can be regarded as sparse foreground anomalies. Then, background structures in slow motions can be treated as the low-rank component. Therefore, RPCA is defined to decompose a given XCA data $D$ into a sum of a vessel and a background component:
\begin{equation}
min\left \| L \right \| _{*} + \lambda \left \| S \right \| _{1}    s.t. D = L + S 
\end{equation} 
{\tiny }where $L$ is the slowly changing background layer, which can be described as a low-rank matrix, and $S$ is the foreground layer, which can be described as a sparse matrix. $\left \|·\right \|_*$ is the nuclear norm (which is the sum of its singular values), $\left \|·\right \|_1$ is the $l_{1}$-norm regularization, and $ \lambda$ is a regularizing parameter to control the detection sensitivity to a number of outliers. The RPCA problem is further expanded into a more general form with the input image $D$ being defined as\cite{solomon2020deep}:
\begin{equation}
D= H_1 L+H_2 S+N
\end{equation} 
where $H_1$ and $H_2$ are the measurement matrices of $L$ and $S$, respectively, representing the physical acquisition mechanisms which are determined by the characteristics of measurement devices (in XCA images, $H_1$ = $H_2$ = $I$), and $N$ is the additive noise. The RPCA is then formulated in a Lagrangian form as:
\begin{equation}
min \frac{1}{2} \left \| M - H_{1}L - H_2S \right \| _{F}^2 + \lambda_1 \left \| L \right \| _{*} +  \lambda_2 \left \| S \right \| _{1,2}
\end{equation}
where $\left \| . \right \| _{1,2}$ is the mixed $l_{1,2}$-norm and $\lambda_1$ and $\lambda_2$ are the regularizing parameters of $L$ and $S$, respectively. The mixed $l_{1,2}$-norm is applied in the cardiovascular extraction task because the structure and position of the vessels change slowly between adjacent frames. We can define this as:
\begin{equation}
X = \begin{bmatrix}
	L\\
	S
\end{bmatrix},
P_1 = \begin{bmatrix}
	I\\
	0
\end{bmatrix},
P_2 = \begin{bmatrix}
	0\\
	I
\end{bmatrix},
A = \begin{bmatrix}
	H_1\\
	H_2
\end{bmatrix}
\end{equation}
Then, Equation (3) can be rewritten as
\begin{equation}
\min_{L,S} \frac{1}{2}  \left \| D-AX \right \| _F ^2 + h(X)
\end{equation}
where $h(X) = \lambda_1 \left \| P_1X \right \| _{*} + \lambda_2 \left \| P_2X \right \| _{1,2}$. Thus, the minimization problem (5) can be regarded as a regularized least-squares problem, which can be solved by the iterative shrinkage/thresholding algorithm, where $L$ and $S$ are iteratively updated until the formula reaches its minimum. $L^{k+1}$ and $S^{k+1}$ at iteration $k + 1$ can be updated\cite{solomon2020deep} via
\begin{equation}
L^{k + 1} = SVT_{\lambda_1 / L_f}{(I - \frac{1}{L_f}H_{1}^{H}H_1)L^k - H_{1}^{H}H_2S^k + H_1^HD}
\end{equation}
\begin{equation}
S^{k + 1} = \psi_{\lambda_2 / L_f}{(I - \frac{1}{L_f}H_{2}^{H}H_2)L^k - H_{2}^{H}H_1S^k + H_2^HD}
\end{equation}
where $SVT_{\lambda_1/L_f}$ is the singular-value thresholding operator, $\psi_{\lambda_2/L_f}$ is the soft-thresholding operator, and $L_f$ is the Lipschitz constant.

\subsection{RPCA Unrolling Network}
Traditional iterative algorithm can be unfolded into a deep neural network, where each layer of the network is represented as one iteration of the algorithm\cite{monga2021algorithm}. Thus, passing through the layers of the deep unfolded network can be viewed as calculation via the iterative algorithm a finite number of times. Following the principle of deep unfolding, the matrices dependent on $H_1$ and $H_2$ in equations (6) and (7) can be replaced with convolutional kernels. We form a deep network by employing convolutional layers $P_1$, $P_2$,..., $P_6$ to replace the matrices dependent on $H_1$ and $H_2$. Convolutional layers are applied rather than fully connected layers, aimed at reducing the number of learned parameters to improve the time and space efficiency of the network. Then, the equations for the $k$th layer in the unfolded network are computed as follows:
\begin{equation}
	L^{k + 1} = SVT_{\lambda_1^k}{P_5^k*L^k+ P_3^k*S^k+P_1^k*D}
\end{equation}	
\begin{equation}
	S^{k + 1} = \psi_{\lambda_2 ^k}{P_6^k*S^k+ P_4^k*L^k+P_2^k*D}
\end{equation}
where $*$ denotes a convolutional operator. The diagram of a single layer of the unfolded network is shown in Fig. 1. Here, the convolutional layers $P_1^k$,…, $P_6^k$, regularization parameters, and $\lambda_1^k$ and $\lambda_2^k$ are learned during the training process.

\subsection{Patch-wise Super-resolution Module}
Unrolling RPCA directly to extract XCA vessels faces limitations since it ignores the influence of additive noise $N$, in which a dynamic background with complex variations and mixed Gaussian-Poisson noise in XCA images largely affects the foreground/background decomposition. Although the regularization parameters $\lambda_1^k$ and $\lambda_2^k$ can be changed to adjust the number of foreground components in the final result to reduce noise to some extent, it is very difficult for the unfolded and traditional RPCA methods to eliminate noisy background disturbances while simultaneously preserving the entire vessel structure, especially the distal vessels.

Considering that SR network can effectively extract structural features of target object and selectively enhance these features without introducing much noise for image segmentation\cite{wang2020dual}, we assume that this SR network's ability to select features can be explored for our work. In addition, inspired by the fact that the mixed Gaussian-Poisson noise locally corrupts the detailed information of vessel branches and can be successfully removed in a patch-wise Gaussian denoising\cite{Irrera2016flexible,zhao2019texture}, we propose a patch-wise SR module with sparse feature selection in RPCA-UNet to extract vessels and eliminate artefacts simultaneously.

The patch-wise SR module is embedded into each iteration/layer of RPCA-UNet to gradually refine vessel extraction and simultaneously remove background artefacts. Inspired by \cite{singh2020eds}, we introduce a feature pooling layer at the beginning of each iteration to downsample the input signal first, which can reduce the influence of redundant information. In RPCA-UNet, the motion artefacts and complex Gaussian-Poisson noise, which are described as $N$ in equation (2), can then be locally discarded to a large extent such that the input of the unfolded RPCA algorithm can be approximately regarded as consisting of only sparse components and low-rank components. 

At the end of each iteration, the residual module and CLSTM network are introduced to iteratively learn the high-level spatiotemporal semantic information of sparsely distributed vessels and refine the output of the deep unfolded RPCA layer. Specifically, the residual module is first applied to extract deep features through multiple convolutional layers. After this feature extraction, the extracted features along with the original features are transferred to the next step via the residual operation. Then, the CLSTM network is applied to combine high-level spatiotemporal semantic details in the whole XCA sequence and selectively adjust the weights of these features. 

Different from traditional LSTM network, which uses full connections, CLSTM network replaces the multiplication operations with convolutional layers. This modification enables the CLSTM network to propagate spatiotemporal features in the training process of deep network\cite{shi2015convolutional}. The key CLSTM network formulation is: 
\begin{align}
	& i_{t} = \sigma  (W_{xi} * X_{t} + W_{hi} * h_{t-1} + W_{ci}\circ c_{t-1} + b_{i}) \notag \\
	& f_{t} = \sigma  (W_{xf} * X_{t} + W_{hf} * h_{t-1} + W_{cf}\circ c_{t-1} + b_{f}) \\	
	&c_t = f_t \circ  c_{t-1} + i_t \circ  tanh(W_{xc} * X_t +W_{hc} * h_{t-1} + b_c) \notag \\
	&o_t = \sigma (W_{xo} * X_t +W_{ho} * h_{t-1} +W_{co} \circ  c_t + b_o) \notag \\
	&h_t = o_t \circ  tanh(c_t) \notag
\end{align}
where * denotes the convolutional operator and $\circ $ denotes the Hadamard product. The memory cell $c_t$ can be used to store the spatiotemporal information of previous frames. The information in the memory cell can be written, cleared and propagated by controlling gates $i_t$, $f_t$, and $o_t$, respectively. The gate $o_t$ that is regarded as a selector can select features from the complementary spatiotemporal information of previous frames to enhance deep features. Then, $h_t$ is the final output, which is determined by the current input and the spatiotemporal information in the memory cell to aid sparse feature selection for better prediction. 

Usually, the CLSTM network can be inserted at different stages of the SR module, such as at the beginning of the module, at the end of the module or during the feature extraction of the residual module. We choose to embed the CLSTM network in the feature extraction such that the weights of extracted features can be selectively adjusted through the spatiotemporal information in memory cells. This embedded feature selection via CLSTM network is assumed to be the core mechanism that enables the sparse feature selection for patch-wise SR vessel extraction and non-vessel artefact removal in RPCA-UNet.

Finally, the output of the CLSTM network is transferred to a sub-pixel convolution layer, which is often used in the image SR task to upscale the output for enhancing detailed information, such as distal and branch vessels in an XCA sequence.
\begin{figure*}[htbp]
	\centering
	\subfigure{
		\begin{minipage}[t]{0.130\linewidth}
			\centering
			\includegraphics[width=1in]{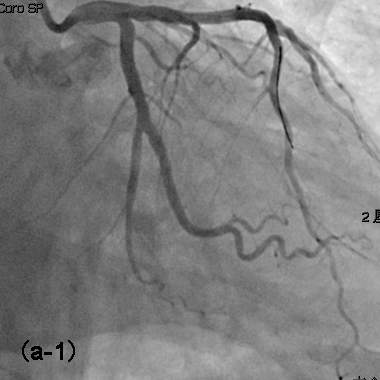}
		\end{minipage}%
	}%
	\subfigure{
		\begin{minipage}[t]{0.13\linewidth}
			\centering
			\includegraphics[width=1in]{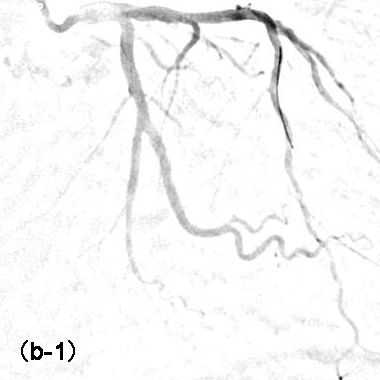}
		\end{minipage}%
	}%
	\subfigure{
		\begin{minipage}[t]{0.13\linewidth}
			\centering
			\includegraphics[width=1in]{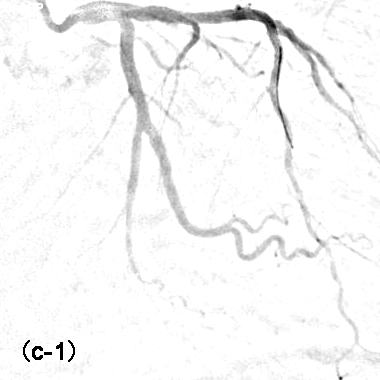}
		\end{minipage}
	}%
	\subfigure{
		\begin{minipage}[t]{0.13\linewidth}
			\centering
			\includegraphics[width=1in]{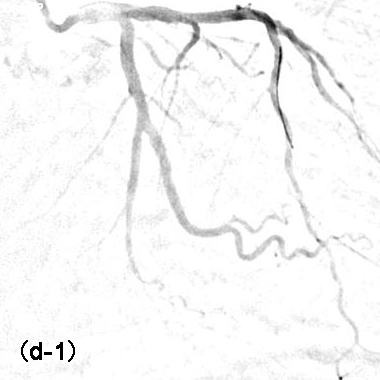}
		\end{minipage}
	}%
	\subfigure{
		\begin{minipage}[t]{0.13\linewidth}
			\centering
			\includegraphics[width=1in]{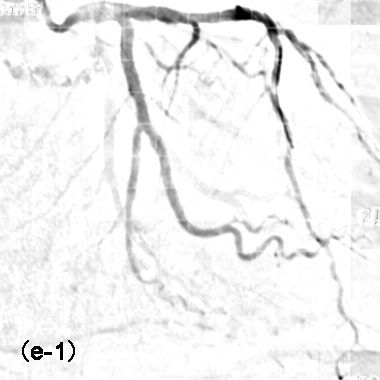}
		\end{minipage}
	}%
	\subfigure{
		\begin{minipage}[t]{0.13\linewidth}
			\centering
			\includegraphics[width=1in]{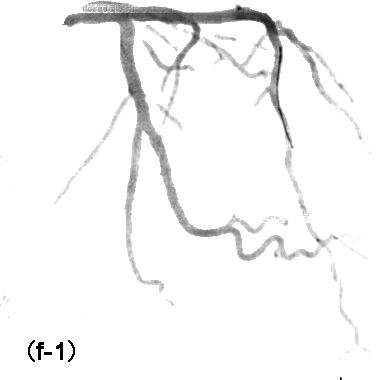}
		\end{minipage}
	}%
	\subfigure{
		\begin{minipage}[t]{0.13\linewidth}
			\centering
			\includegraphics[width=1in]{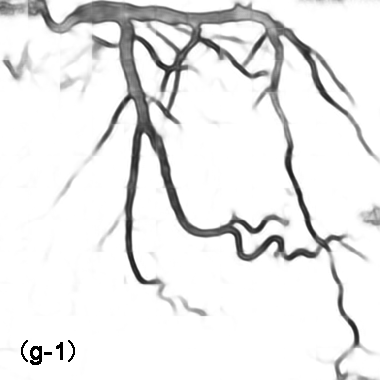}
		\end{minipage}
	}%
	
	\subfigure{
		\begin{minipage}[t]{0.13\linewidth}
			\centering
			\includegraphics[width=1in]{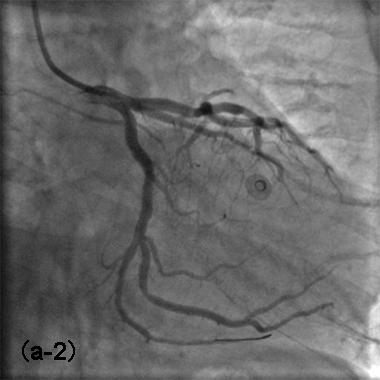}
		\end{minipage}%
	}%
	\subfigure{
		\begin{minipage}[t]{0.13\linewidth}
			\centering
			\includegraphics[width=1in]{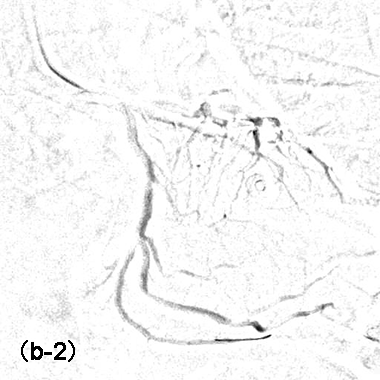}
		\end{minipage}
	}%
	\subfigure{
		\begin{minipage}[t]{0.13\linewidth}
			\centering
			\includegraphics[width=1in]{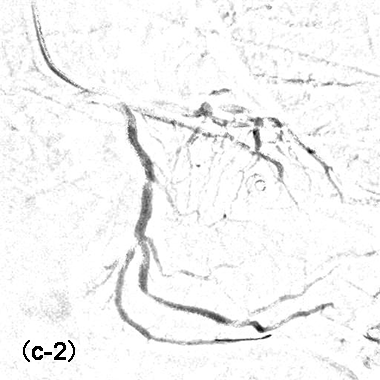}
		\end{minipage}
	}%
	\subfigure{
		\begin{minipage}[t]{0.13\linewidth}
			\centering
			\includegraphics[width=1in]{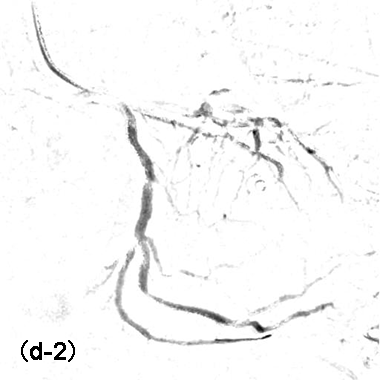}
		\end{minipage}
	}%
	\subfigure{
		\begin{minipage}[t]{0.13\linewidth}
			\centering
			\includegraphics[width=1in]{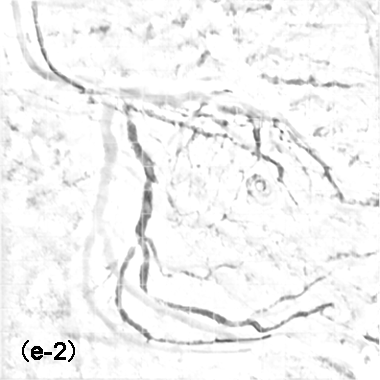}
		\end{minipage}
	}%
	\subfigure{
		\begin{minipage}[t]{0.13\linewidth}
			\centering
			\includegraphics[width=1in]{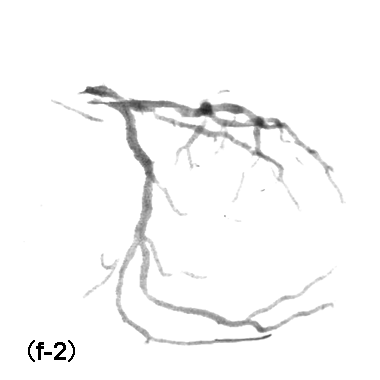}
		\end{minipage}
	}%
	\subfigure{
		\begin{minipage}[t]{0.13\linewidth}
			\centering
			\includegraphics[width=1in]{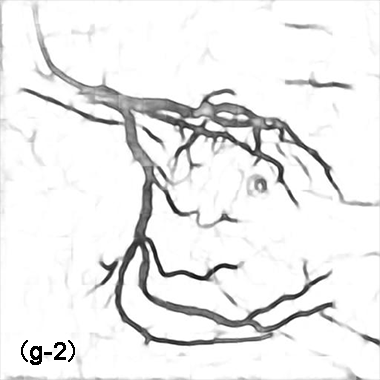}
		\end{minipage}
	}%
	
	\subfigure{
		\begin{minipage}[t]{0.13\linewidth}
			\centering
			\includegraphics[width=1in]{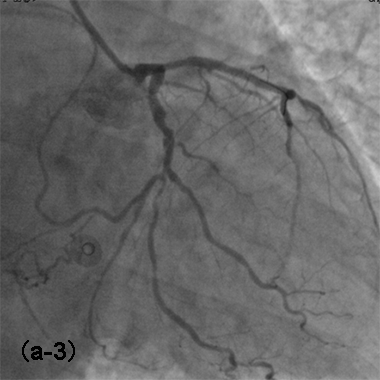}
		\end{minipage}%
	}%
	\subfigure{
		\begin{minipage}[t]{0.13\linewidth}
			\centering
			\includegraphics[width=1in]{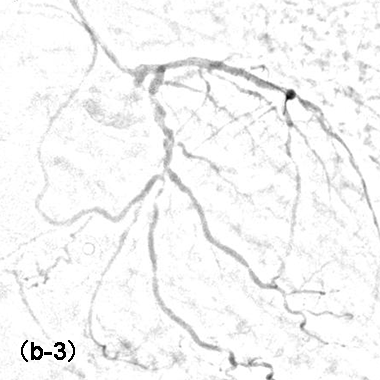}
		\end{minipage}
	}%
	\subfigure{
		\begin{minipage}[t]{0.13\linewidth}
			\centering
			\includegraphics[width=1in]{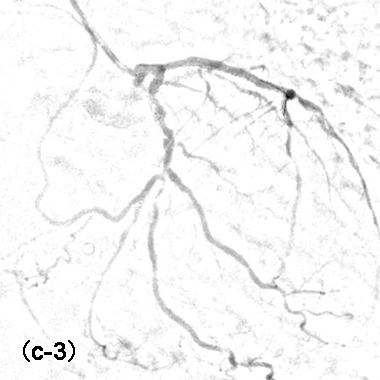}
		\end{minipage}
	}%
	\subfigure{
		\begin{minipage}[t]{0.13\linewidth}
			\centering
			\includegraphics[width=1in]{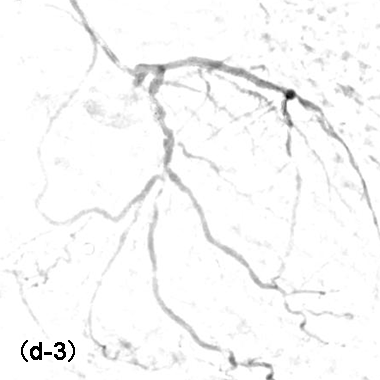}
		\end{minipage}
	}%
	\subfigure{
		\begin{minipage}[t]{0.13\linewidth}
			\centering
			\includegraphics[width=1in]{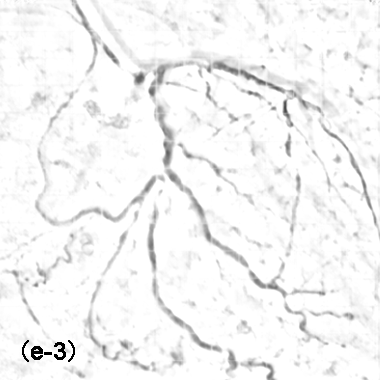}
		\end{minipage}
	}%
	\subfigure{
		\begin{minipage}[t]{0.13\linewidth}
			\centering
			\includegraphics[width=1in]{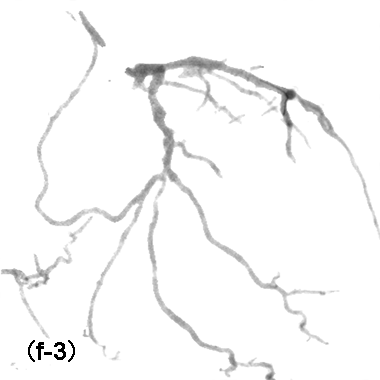}
		\end{minipage}
	}%
	\subfigure{
		\begin{minipage}[t]{0.13\linewidth}
			\centering
			\includegraphics[width=1in]{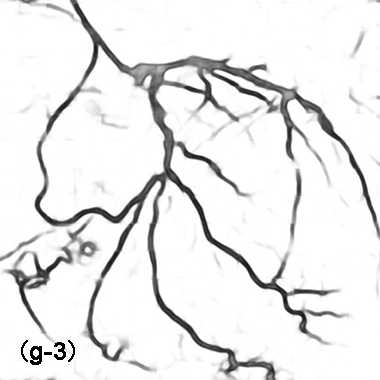}
		\end{minipage}
	}%
	\centering	
	\caption{XCA vessel extraction results. (a) Original XCA image; (b) ALF-RPCA; (c) MoG-RPCA; (d) MCR-RPCA\cite{2017Extracting}; (e) CORONA\cite{solomon2020deep}; (f) VRBC\cite{qin2019accurate}; (g) RPCA-UNet.}	
\end{figure*}

\subsection{Automatic Vessel Labelling}
RPCA-UNet aims to extract both the geometrical features and image grey values of XCA vessels, which are unrealistic to be labelled manually for the deep learning of RPCA-UNet. Therefore, RPCA-UNet implemented weakly supervised learning through an automatic vessel labelling with tensor-completion-based VRBC method \cite{qin2019accurate}, which is the only method available to accurately and automatically recover vessel intensity information with rarely introducing background component. Specifically, VRBC first extracts vessel structures from complex and noisy backgrounds by RPCA-based vessel extraction. An accurate binary mask of vessel is then finely generated via Radon-like feature filtering with spatially adaptive thresholding. Subsequently, vessel-masked background regions are recovered to complete background layers by implementing tensor completion with the spatiotemporal consistency of whole background regions. Finally, the layers containing vessels' greyscale values can be accurately extracted by subtracting the completed background layers from the overall XCA images.


\section{Experimental Results}
\subsection{Experimental Materials}
Our experiments collected 43 sequences of real clinical XCA images from Renji Hospital of Shanghai Jiao Tong University. The length of each sequence ranges from 30 to 140 frames. Images from these XCA sequences were manually annotated by three experts to obtain the vessel mask ground truth for evaluation. To eliminate differences in size, these frames were resized to 512 $\times$ 512 resolution with 8 bits per pixel. It is worth noting that these sequences are heterogeneous since they are collected from different machines, including a medical angiography X-ray system from Philips and the 800 mAh digital silhouette angiography X-ray machine from Siemens. Therefore, the noise distribution and the pixel grey level range of each sequence are very different. 

\subsection{Experiment Settings and RPCA-UNet Training}
RPCA-UNet consists of 4 layers. The first two layers use convolutional kernels of size = 5 with stride = 1, padding = 2 and a bias, and the other two layers use convolutional kernels of size = 3 with stride = 1, padding = 1 and a bias. We choose the ADAM optimizer with a learning rate of 0.0001. In the feature selection module, the average pooling layer with pooling window = 2 and stride = 2 is selected. The upscaling rate of the SR module is set to 2.

RPCA-UNet is trained using back-propagation in a weakly-supervised manner. Training pairs of vessel/background labelling are generated by the VRBC method\cite{qin2019accurate} and the training images are divided into 64 $\times$ 64 $\times$ 20 patches with a certain overlap (50\% between two neighboring patches). A total of 15 sequences containing 900 samples are used in the experiment and the total amount of patches used in dataset is 20000. Then, the dataset is randomly divided into training, validation, and test datasets at a ratio of approximately 0.6:0.2:0.2, respectively. The patches in the output are spliced with their grey values being the average of overlapping patches. The loss function is chosen as the sum of the mean square errors between the predicted vessel/background values of the network and the corresponding vessel/background labels.

\subsection{Comparison Methods}
We used the VRBC\cite{qin2019accurate} and several state-of-the-art RPCA-based methods for comparison, which include ALF-RPCA \cite{wang2018robust}, MoG-RPCA \cite{zhao2014robust}, our previous MCR-RPCA \cite{2017Extracting} and CORONA \cite{solomon2020deep}. Moreover, to evaluate the performance of our network on the vessel segmentation task, we compared the vessel segmentation results with those of several other vessel segmentation algorithms, including Coye's\cite{coye2015novel}, Frangi's\cite{frangi1998multiscale} results and those of the deep-learning-based SVS-net\cite{hao2020sequential} and $\text{CS}^2$-Net \cite{mou2021cs2}. The parameters of these segmentation algorithms were tuned to achieve the best performance.

\subsection{Visual Evaluation on Experimental Results}
Vessel extraction results are shown in {Fig. 2}, in which three RPCA-based methods can extract major vessels relatively well but obvious vessel-like residuals with considerable noises still remain (see Fig. 2(b)-(d)). Moreover, distal vessels are hardly extracted because they are completely submerged in the noisy background. Although CORONA performs better in distal vessel extraction (see Fig. 2(e)), the noisy background still has a large influence on the extraction results, as for the traditional RPCA-based methods. The VRBC framework extracts much better grey value vessels than the above methods, with most of the background artefacts being removed (see Fig. 2(f)). However, the performance of VRBC in the extraction of vessel branches and distal vessels is still not satisfactory. Compared to these methods, RPCA-UNet greatly improves the vessel extraction performances since the extracted vessel tree structure is clearer and more complete, especially for the vessel branches and distal vessels (see Fig. 2(g)).

It is worth noting that the visual contrast of vessel extraction (see Fig. 2) of RPCA-UNet is obviously enhanced compared with that of the VRBC method \cite{qin2019accurate}. Specifically, in the recovered profiles of vessel intensity, the grey level of distal vessels is lower and that of large vessels is increasingly higher. These results are entirely consistent with the X-ray attenuation coefficients of various structures imaged in the XCA images. During low-dose XCA imaging, the amount of contrast agent in large vessels is greater than that of distal vessels, which makes the grey level of large vessels higher than that of distal vessels. Therefore, the recovered grey levels of vessels achieved by RPCA-UNet is in high fidelity relative to the real distribution of contrast agent in XCA vessels, which is helpful for quantitatively analysing the structure-functional characterization of cardiac perfusion.

In vessel segmentation evaluation, we use different colours to label the pixels of segmentation results by comparing with ground truth vessel mask (see Fig. 3(b)), in which green pixels represent the true positive pixels that are correctly classified as vessels, blue pixels represent false negative pixels that are vessel pixels but wrongly classified as backgrounds, red pixels are false positive pixels that are wrongly classified as vessels but practically belonging to the backgrounds. The segmentation results in Fig. 3 show that Coye's and Frangi's methods detect either too few vessels or too much noises (see Fig. 3(c)-(d)). These traditional methods have poor performance in the foreground and background areas with similar grey values. SVS-net can detect most of the vessels and effectively suppress the background noise (see Fig. 3(e)). However, in some distal vessel areas, the detection result appears to have missing or discontinuous segments. $\text{CS}^2$-Net has a great segmentation performance on distal vessels while in some cases it may introduce vessel-like noisy artefacts (see Fig. 3(f)). Moreover, its comprehensive performance seems to be unstable since discontinuities appear in some major blood vessels. RPCA-UNet performs best in the segmentation experiments since almost all the major vessels and most of the distal vessels can be detected (see Fig. 3(g)). Although some background residue still exists in the detection results, it does not have a significant impact on the visual effect.
\begin{figure*}[htbp]
	\centering
	\subfigure{
		\begin{minipage}[t]{0.13\linewidth}
			\centering
			\includegraphics[width=4in]{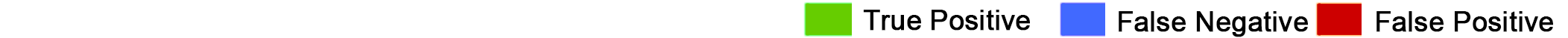}
		\end{minipage}%
	}%
	
	\subfigure{
		\begin{minipage}[t]{0.132\linewidth}
			\centering
			\includegraphics[width=1in]{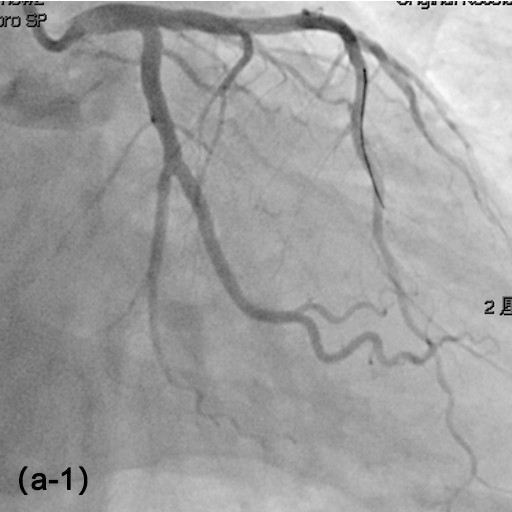}
		\end{minipage}%
	}%
	\subfigure{
		\begin{minipage}[t]{0.133\linewidth}
			\centering
			\includegraphics[width=1in]{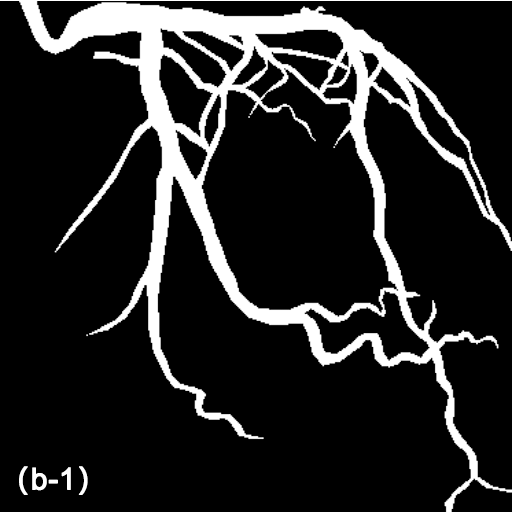}
		\end{minipage}%
	}%
	\subfigure{
		\begin{minipage}[t]{0.129\linewidth}
			\centering
			\includegraphics[width=1in]{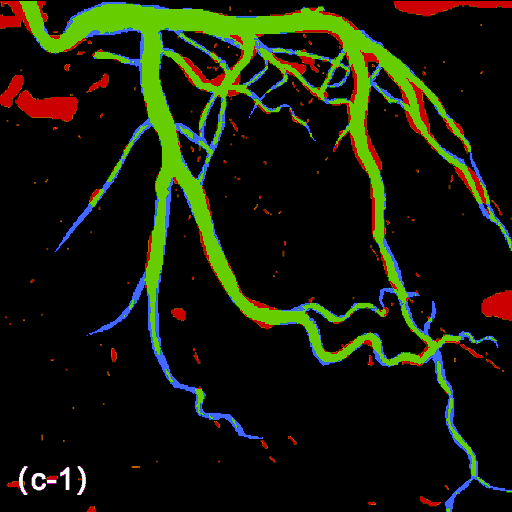}
		\end{minipage}
	}%
	\subfigure{
		\begin{minipage}[t]{0.129\linewidth}
			\centering
			\includegraphics[width=1in]{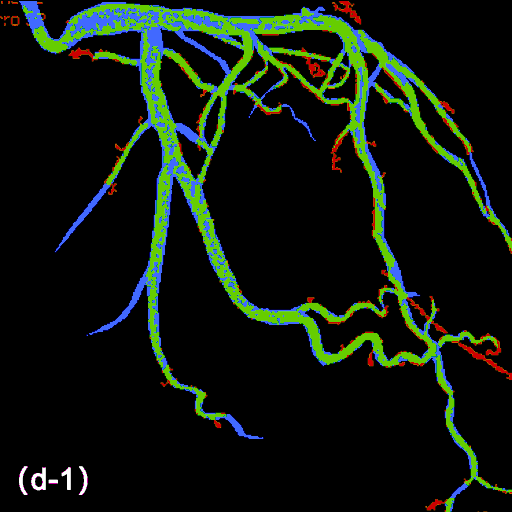}
		\end{minipage}
	}%
	\subfigure{
		\begin{minipage}[t]{0.128\linewidth}
			\centering
			\includegraphics[width=1in]{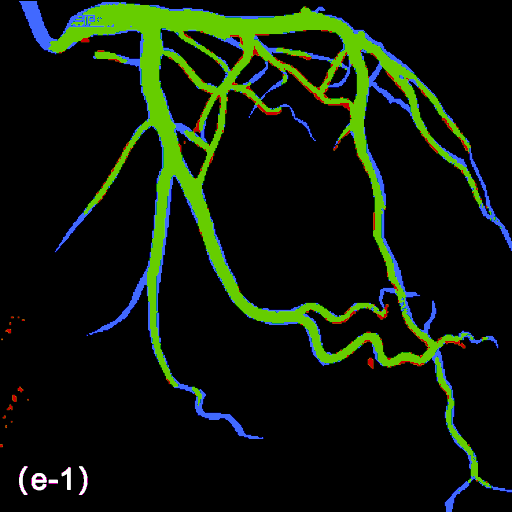}
		\end{minipage}
	}%
	\subfigure{
		\begin{minipage}[t]{0.129\linewidth}
			\centering
			\includegraphics[width=1in]{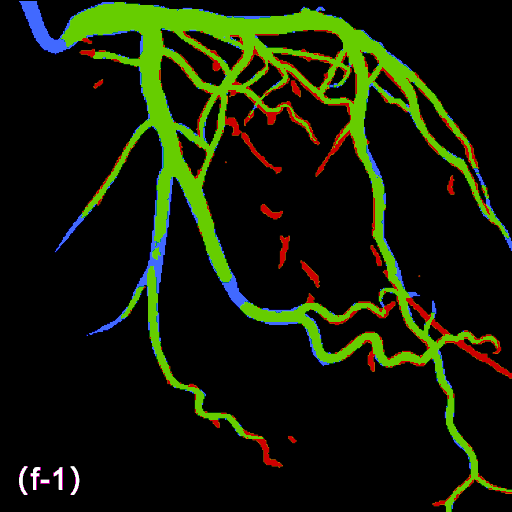}
		\end{minipage}
	}%
	\subfigure{
		\begin{minipage}[t]{0.129\linewidth}
			\centering
			\includegraphics[width=1in]{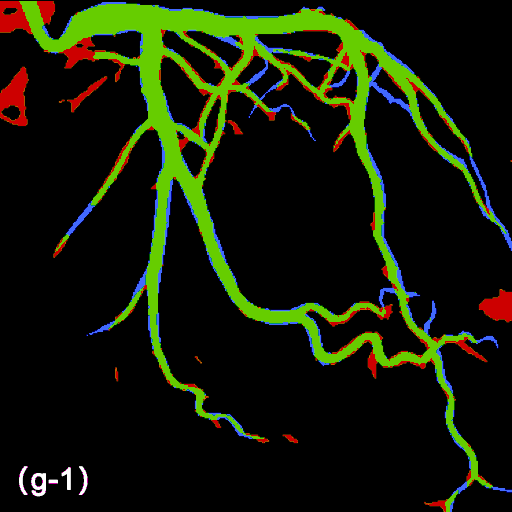}
		\end{minipage}
	}%
	
	\subfigure{
		\begin{minipage}[t]{0.133\linewidth}
			\centering
			\includegraphics[width=1in]{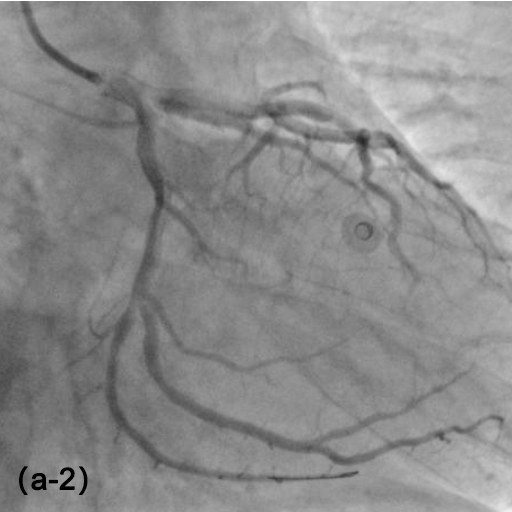}
		\end{minipage}%
	}%
	\subfigure{
		\begin{minipage}[t]{0.133\linewidth}
			\centering
			\includegraphics[width=1in]{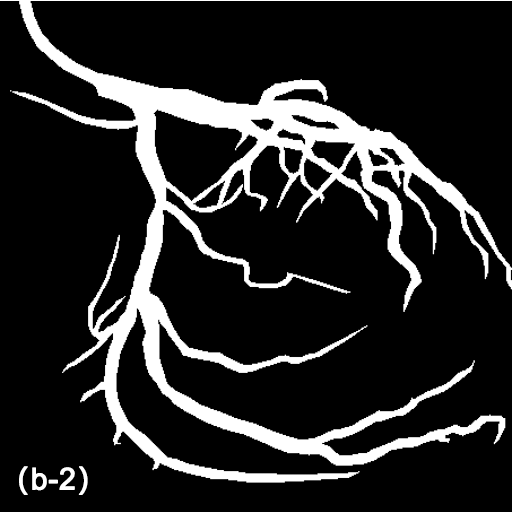}
		\end{minipage}%
	}%
	\subfigure{
		\begin{minipage}[t]{0.129\linewidth}
			\centering
			\includegraphics[width=1in]{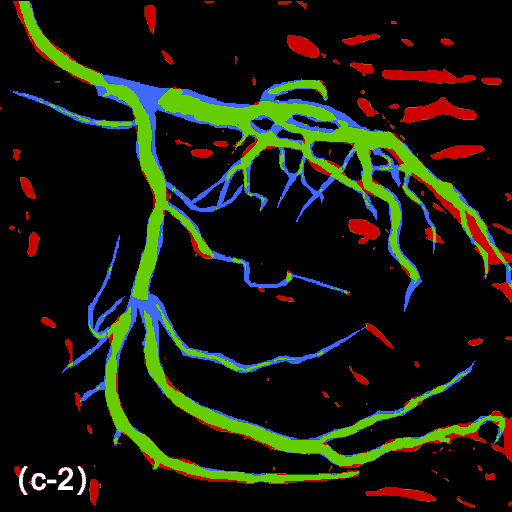}
		\end{minipage}
	}%
	\subfigure{
		\begin{minipage}[t]{0.129\linewidth}
			\centering
			\includegraphics[width=1in]{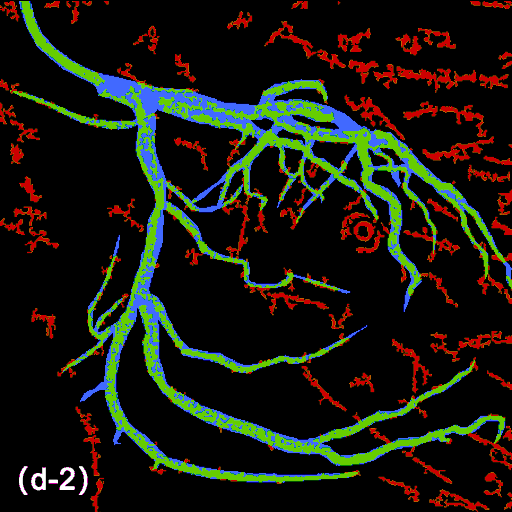}
		\end{minipage}
	}%
	\subfigure{
		\begin{minipage}[t]{0.129\linewidth}
			\centering
			\includegraphics[width=1in]{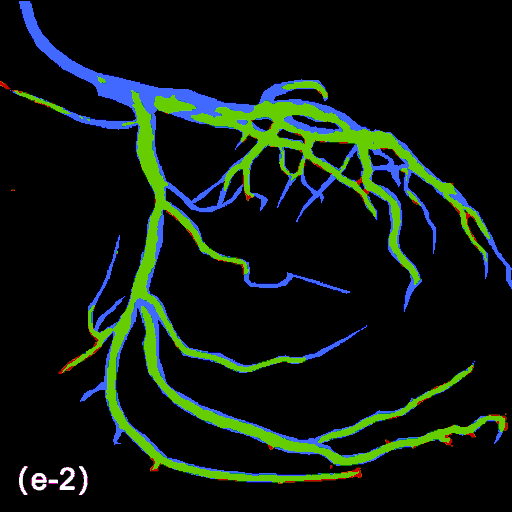}
		\end{minipage}
	}%
	\subfigure{
		\begin{minipage}[t]{0.129\linewidth}
			\centering
			\includegraphics[width=1in]{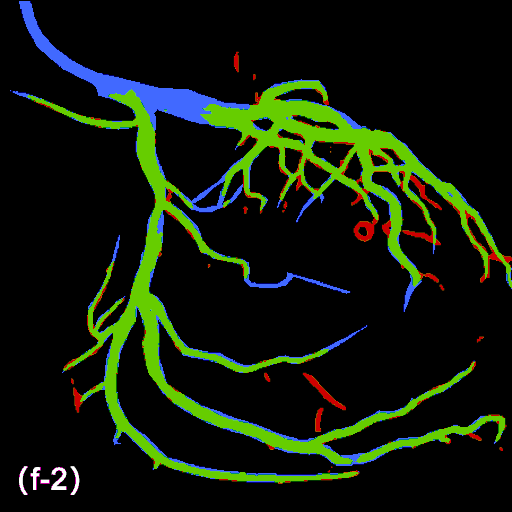}
		\end{minipage}
	}%
	\subfigure{
		\begin{minipage}[t]{0.129\linewidth}
			\centering
			\includegraphics[width=1in]{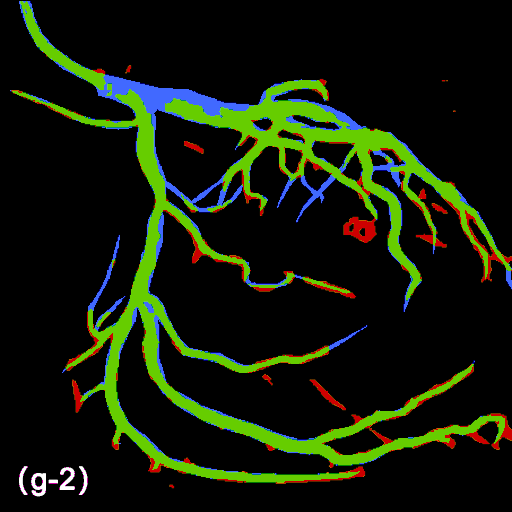}
		\end{minipage}
	}%
	
	\subfigure{
		\begin{minipage}[t]{0.133\linewidth}
			\centering
			\includegraphics[width=1in]{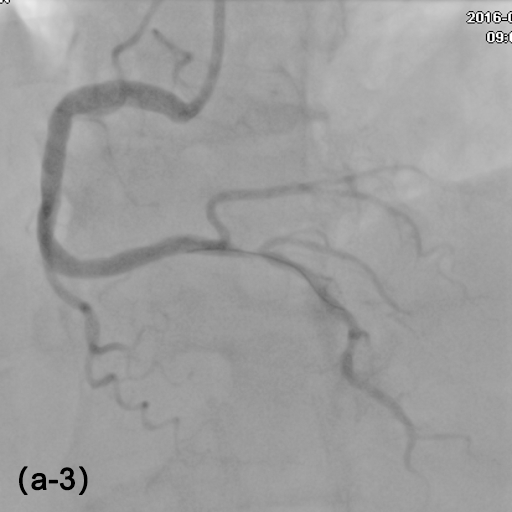}
		\end{minipage}%
	}%
	\subfigure{
		\begin{minipage}[t]{0.133\linewidth}
			\centering
			\includegraphics[width=1in]{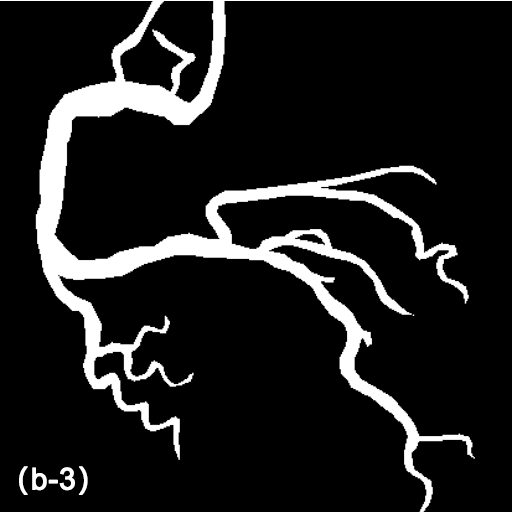}
		\end{minipage}%
	}%
	\subfigure{
		\begin{minipage}[t]{0.129\linewidth}
			\centering
			\includegraphics[width=1in]{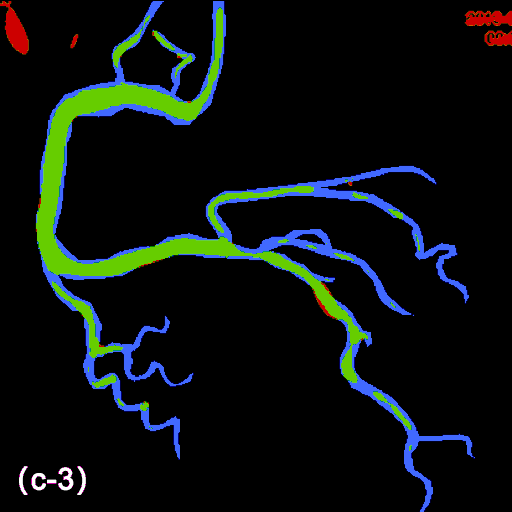}
		\end{minipage}
	}%
	\subfigure{
		\begin{minipage}[t]{0.129\linewidth}
			\centering
			\includegraphics[width=1in]{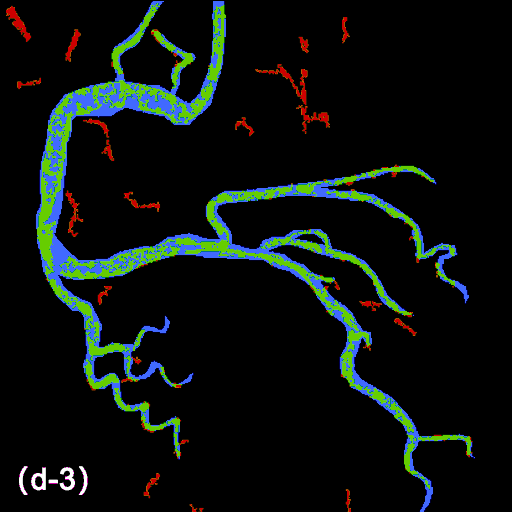}
		\end{minipage}
	}%
	\subfigure{
		\begin{minipage}[t]{0.129\linewidth}
			\centering
			\includegraphics[width=1in]{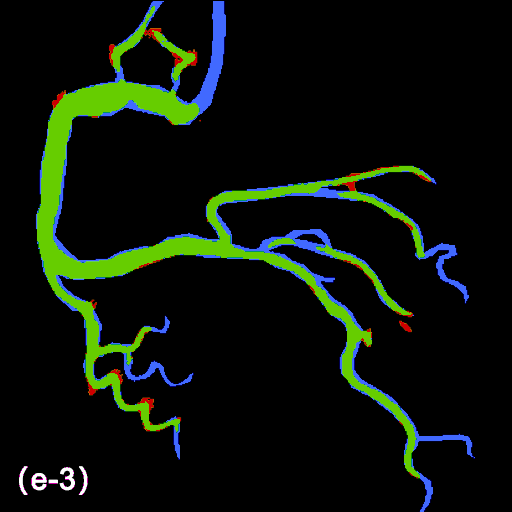}
		\end{minipage}
	}%
	\subfigure{
		\begin{minipage}[t]{0.129\linewidth}
			\centering
			\includegraphics[width=1in]{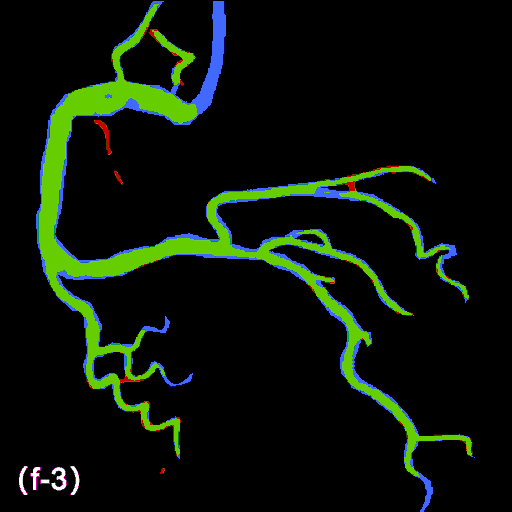}
		\end{minipage}
	}%
	\subfigure{
		\begin{minipage}[t]{0.129\linewidth}
			\centering
			\includegraphics[width=1in]{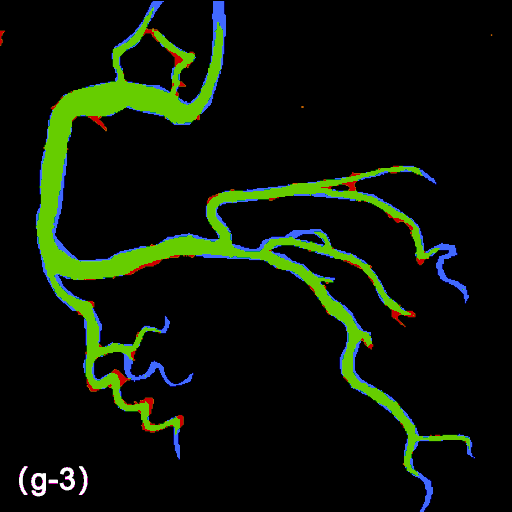}
		\end{minipage}
	}%
	
	\caption{XCA vessel segmentation results. Pixels labelled with green, blue, and red colours represent true positive pixels, false negative pixels, and false positive pixels, respectively. (a) Original XCA image; (b) Ground-truth vessel mask; (c) Frangi's; (d) Coye's; (e) SVS-net; (f) $\text{CS}^2$-Net; (g) RPCA-UNet.}
\end{figure*}

\subsection{Quantitative Evaluation of Vessel Extraction}
The vessel visibility can be quantitatively evaluated by using the contrast-to-noise ratio (CNR)\cite{solomon2020deep} of the vessel layer images. The CNR measures the contrast between the vessels and backgrounds, where a larger CNR value means a better vessel visibility. The CNR can be calculated by:
\begin{equation}
	CNR =  \frac{\left | \mu_{V} - \mu_{B}  \right |}{ \sqrt{\sigma_B^{2} + \sigma_V^{2}}  } 	
\end{equation}
where $\mu_{V}$ and $\mu_{B}$ are the pixel intensity means in the vessel and background regions, respectively, and $\sigma_V$ and $\sigma_B$ are the standard deviation of the pixel intensity values in the vessel regions and background regions, respectively.

To further evaluate vessel visibility, we define global and local background regions to cover all the image regions except the vessel regions and the 7-pixel-wide neighbourhood regions surrounding the vessel regions, respectively. The CNR calculation results are shown in \tablename { I} and Fig. 4. The results show that RPCA-UNet achieves the highest global and local CNRs, which indicates that RPCA-UNet greatly improves the vessel visibility both globally and locally because it achieves excellent vessel extraction and noise suppression. 

To evaluate the time efficiency of RPCA-UNet, we calculate the average running time per image of the above vessel extraction methods and the results are shown in \tablename { I}. RPCA-UNet has relatively fast speed even though it divides each image into patches with 50\% overlapping to eliminate the influence of mixed noise, which will increase the amount of calculation by about 4 times. Such sacrifice in speed for better extraction results is assumed to be worthwhile in meeting the clinical demand. Moreover, the parameter sizes of RPCA-UNet and SVS-net is 0.76M and 10.06M, respectively. This indicates that the parameter size of RPCA-UNet is quite small and its storage efficiency is remarkable.
\begin{table}\normalsize
	\caption{Performance of different vessel extraction methods in terms of CNR values (mean$\pm$ standard deviation)}
	\label{1}
	\setlength{\tabcolsep}{2.7mm}{
		\begin{tabular}{llll}			
			\hline			
			Method& Global CNR& Local CNR & Time(s)\\ 
			\hline
			MCR-RPCA& 1.01 $\pm$ 0.19& 1.00 $\pm$ 0.20 & 23.19\\
			MoG-RPCA& 1.06 $\pm$ 0.22& 1.06 $\pm$ 0.21 & 1.03\\
			ALF-RPCA& 0.93 $\pm$ 0.25& 0.95 $\pm$ 0.24 & 0.20\\
			CORONA& 0.96 $\pm$ 0.18& 1.01 $\pm$ 0.183 & 0.61\\
			VRBC& 1.04 $\pm$ 0.14& 1.02 $\pm$ 0.14 & 24.68\\
			RPCA-UNet & \textbf{1.78 $\pm$ 0.25}& \textbf{1.65 $\pm$ 0.20} & 0.92 \\
			\hline
	\end{tabular}}
\end{table}
\begin{figure}[htbp] 
	\centering 
	\includegraphics[width=0.48\textwidth]{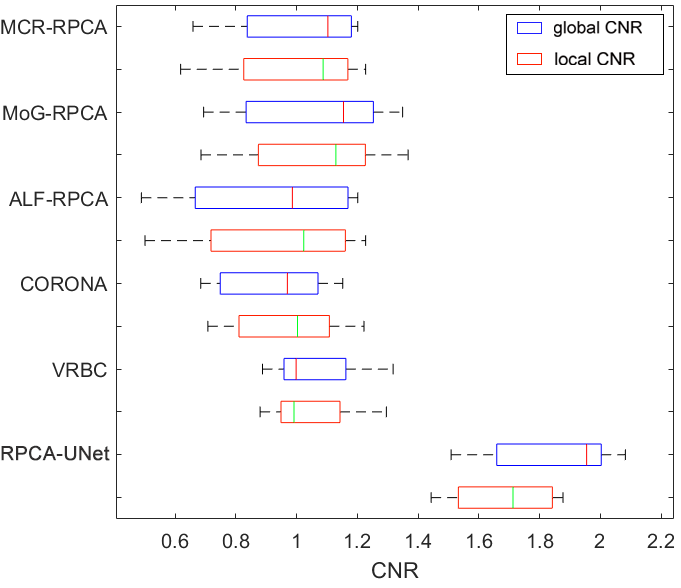}
	\caption{Performance of vessel extraction methods in terms of global and local CNRs for twelve real XCA images}  
	\label{4}
\end{figure}

\subsection{Quantitative Evaluation of Vessel Segmentation}
To evaluate the performances of the segmentation results of the proposed method, the detection rate (DR), precision (P), and F-measure (F) are employed. We also compare RPCA-UNet with other state-of-the-art methods. The abovementioned metrics can be calculated as follows:
\begin{equation}
	 DR = \frac{TP}{TP + FN} , P = \frac{TP}{TP + FP} , F = \frac{2\times DR \times P}{DR + P} 
\end{equation}
where TP is the total number of true positive pixels, FP indicates the total number of  false positive pixels, and TN and FN indicate the total numbers of true negative and false negative pixels that are correctly classified as background pixels and wrongly predicted as background pixels in the segmentation output, respectively. The DR represents the proportion between the correctly classified vessel pixels and the total vessel pixels in the ground truth, P represents the ratio of the TP among all TP, and F comprehensively considers both the P and DR metrics and indicates the overall segmentation performance. All these metrics range from 0 to 1, where higher values mean a better segmentation performance.


In the quantitative evaluation of the segmentation experiments, 12 images selected from different sequences are manually annotated as the ground truth. The DR, P, and F-measure of these 12 images are measured and displayed in \tablename{ II}. RPCA-UNet generally obtains the highest DR and F scores. RPCA-UNet achieves relatively lower P value than some other methods do, which represents the proportion of positive examples that are actually positive. This is because that RPCA-UNet tends to extract complete vessel information as much as possible and may inevitably cause an increase in false positives, which leads to a relative low P. However, the improvement of vessel  detection makes RPCA-UNet achieve better performance in terms of both DR and F-measure. RPCA-UNet performs best in term of F-measure that represents the comprehensive performance of the P and the DR. Moreover, the XCA images typically selected in our experiments contain many distal vessel branches with a very low contrast to the background components such that it is quite difficult to detect all distal vessels correctly. Therefore, the DR and F results of state-of-the-art methods and RPCA-UNet are generally low. Under these experimental settings, RPCA-UNet still obtained relatively high results due to the larger improvement in its capability to detect vessel branches and distal vessels. Therefore, we believe RPCA-UNet performs better than other methods even if its P is not the highest.
\begin{table} \normalsize
	\caption{Means and Standard Deviations of the DR, P, and F metrics on twelve XCA images}
	\label{2}
	\begin{center}	
		\setlength{\tabcolsep}{1.0mm}{
			\begin{tabular}{lccc}				
				\hline				
				Method& Detection Rate& Precision & F-measure\\
				\hline				
				Coye's& 0.592 $\pm$ 0.087& 0.810 $\pm$ 0.134& 0.675 $\pm$ 0.065      \\
				Frangi's& 0.577 $\pm$ 0.120& 0.686 $\pm$ 0.181&  0.617 $\pm$ 0.126    \\
				SVS-net& 0.635 $\pm$ 0.090& 0.948 $\pm$ 0.030&  0.757 $\pm$ 0.063   \\
				$\text{CS}^2$-Net& 0.747 $\pm$ 0.119& 0.842 $\pm$ 0.089&  0.778 $\pm$ 0.063   \\
				RPCA-UNet& \textbf{0.810 $\pm$ 0.057}& 0.774 $\pm$ 0.161& \textbf{0.783 $\pm$ 0.086}      \\
				\hline
		\end{tabular}}
	\end{center} 
\end{table}

\subsection{Ablation Study}
Several ablation experiments are tested to validate the effectiveness of RPCA-UNet architecture. We compare the vessel extraction results of RPCA-UNet using different iterations/layers with different results from CORONA\cite{solomon2020deep}, SR network, RPCA-UNet with SR module but without CLSTM network. The layers of RPCA-UNet can gradually extract the moving contrast agents and prune complex vessel-like artefacts. As in \cite{solomon2020deep}, for each layer number, we construct RPCA-UNet with that number of layers. These networks are trained for 50 epochs on the same training dataset. 

Fig. 5 shows the comparison results of ablation study. CORONA\cite{solomon2020deep} can extract vessel grey value information while the result (see Fig. 5(b)) is severely disturbed by noisy artefacts. The output of SR network shows clear vessel contour (see Fig. 5(c)) while the grey value information is almost lost since SR network is mainly designed for extracting structure information. The RPCA unrolling network embedded with SR module but without CLSTM network (see Fig. 5(d)) improves the vessel extraction result significantly. However, such combined network has an imperfect performance in extracting distal vessels with low contrast and strong background noises. 
\begin{figure}[htbp]
	\centering
	\subfigure{
		\begin{minipage}[t]{0.22\linewidth}
			\centering
			\includegraphics[width=0.81in]{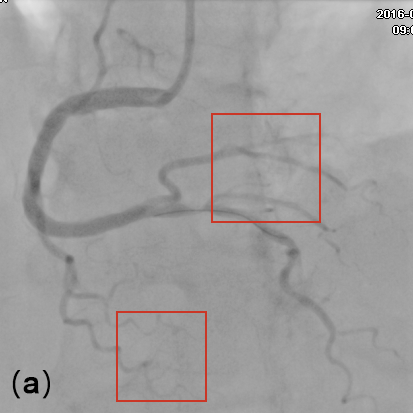}
		\end{minipage}%
	}%
	\subfigure{
		\begin{minipage}[t]{0.22\linewidth}
			\centering
			\includegraphics[width=0.82in]{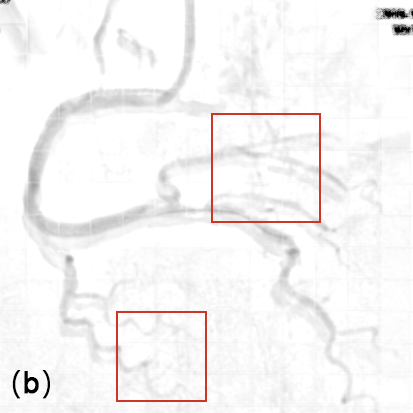}
		\end{minipage}%
	}%
	\subfigure{
		\begin{minipage}[t]{0.22\linewidth}
			\centering
			\includegraphics[width=0.82in]{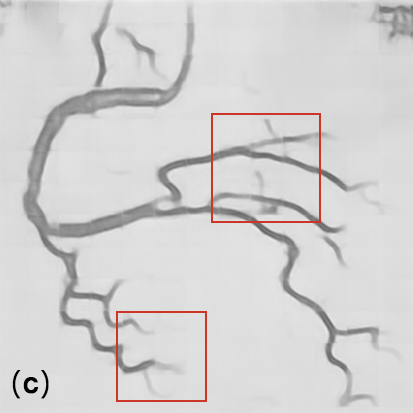}
		\end{minipage}%
	}%
	\subfigure{
		\begin{minipage}[t]{0.22\linewidth}
			\centering
			\includegraphics[width=0.82in]{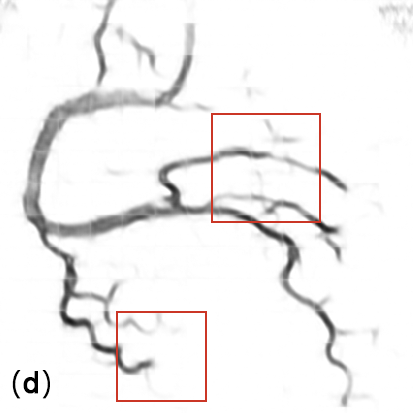}
		\end{minipage}%
	}%
	
	\subfigure{
		\begin{minipage}[t]{0.22\linewidth}
			\centering
			\includegraphics[width=0.82in]{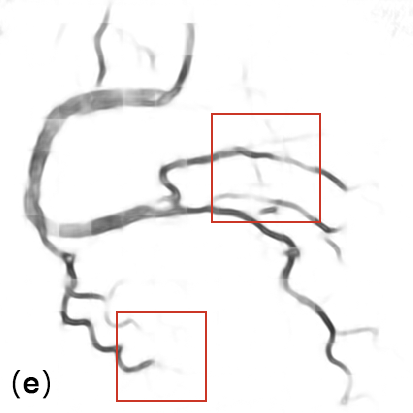}
		\end{minipage}%
	}%
	\subfigure{
		\begin{minipage}[t]{0.22\linewidth}
			\centering
			\includegraphics[width=0.82in]{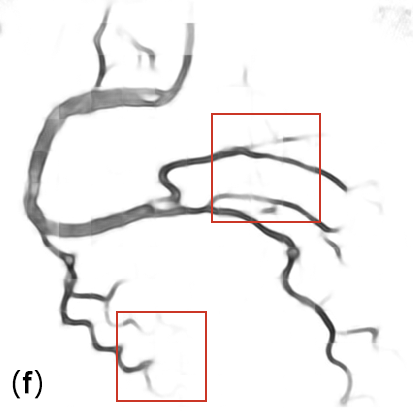}
		\end{minipage}%
	}%
	\subfigure{
		\begin{minipage}[t]{0.22\linewidth}
			\centering
			\includegraphics[width=0.82in]{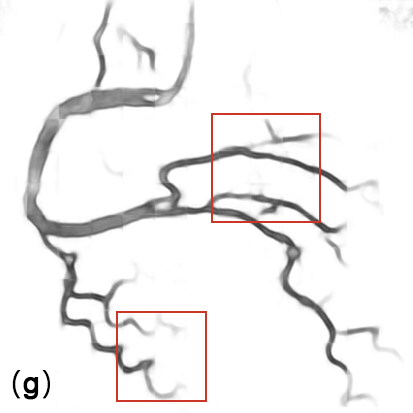}
		\end{minipage}%
	}%
	\subfigure{
		\begin{minipage}[t]{0.22\linewidth}
			\centering
			\includegraphics[width=0.82in]{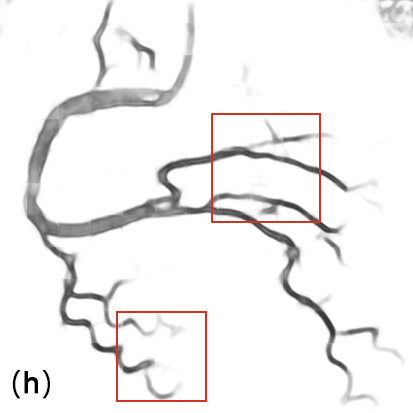}
		\end{minipage}%
	}%
	\caption{The results of ablation study. (a) original XCA image; (b) CORONA\cite{solomon2020deep}; (c) SR network; (d) RPCA-UNet with SR but without CLSTM network; (e)-(h) RPCA-UNet with 1-4 layers.}
\end{figure}

The second row of Fig. 5 shows the comparison between different versions of RPCA-UNet with different layers (see Fig. 5(e)-(h)). The results show that integrating our CLSTM-network-based SR module into RPCA-UNet can extract vessel information iteratively and gradually prune complex vessel-like artefacts and noisy backgrounds. We also compare the mean squared error (MSE) of RPCA-UNet and the results in \tablename{ III} show that the MSE decreases as the number of layers increases. 
\begin{table} \normalsize
	\caption{Mean squared errors of RPCA-UNet with different layers}
	\label{2}
	\begin{center}	
		\setlength{\tabcolsep}{1.0mm}{
			\begin{tabular}{lcccc}				
				\hline				
				Layer Number & 1& 2 & 3 & 4\\
				\hline				
				MSE & 8.29e-03& 6.68e-03 & 6.56e-03 & 6.26e-03      \\
				
				\hline
		\end{tabular}}
	\end{center} 
\end{table}

\subsection{Coarse versus Fine Labels for Weakly Supervised Learning}
Due to our weakly supervised method replacing manual annotation with automatic vessel labelling by VRBC method\cite{qin2019accurate}, it would be interesting to see whether this replacement with automatic vessel labelling had influenced the results. Although obtaining a large number of manually annotated vessel masks with grey values is almost impossible, we have manually annotated binary labels for our proposed SVS-net\cite{hao2020sequential}, whose output is then used as a binary-mask in VRBC method for automatic vessel labelling (refer to the results in Fig. 2(f)). Therefore, we can provide coarse and fine vessel labels to evaluate the impact of labelling quality on the vessel extraction results. 
	
To assess the impact of coarsely labelled versus finely labelled data on weakly supervised vessel extraction, we have automatically generated three types of grey value labels using VRBC method with their corresponding binary masks being first segmented in different ways: original segmentation method\cite{jin2018low} adopted in the VRBC method, SVS-net with training data being generated by the original segmentation method, SVS-net with training data being manually annotated. With these different binary segmentation masks, the grey value labels generated by the VRBC method are displayed in the first row of Fig. 6. We assume that the quality of fine grey value labels generated by the VRBC plus SVS-net with manual annotations is close to that of manual annotations. The vessel extraction results via RPCA-UNet are shown in the second row of Fig. 6. The two networks trained by the first two types of grey value labels similarly achieve great performances while the network trained by the third type of grey value labels, where the labelling quality is the highest and is close to manual annotations, introduces more background impurities in some test cases. We believe the reason for this phenomenon is that too fine labels will cause overfitting in the trained neural network that may have poor generalization ability, thereby erroneously identifying some background components as vessels in some noisy XCA images.
\begin{figure}[htbp]
	\centering	
	\centering 
	\includegraphics[width=0.48\textwidth]{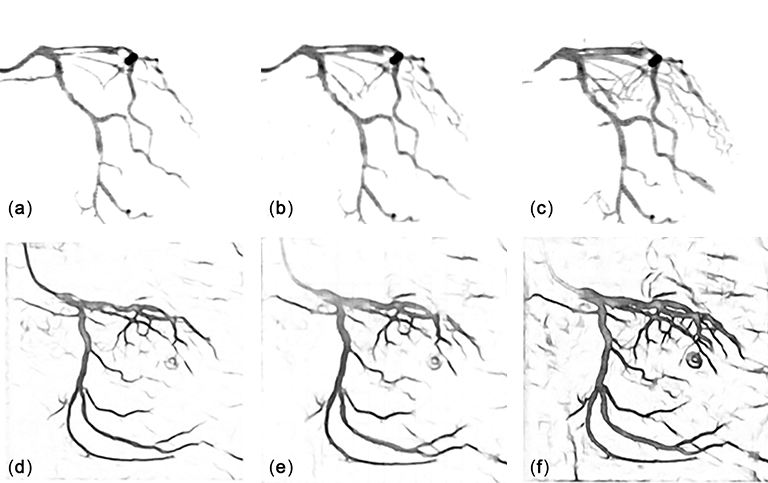}	
	\caption{The effect of coarse versus fine vessel label on the result of weakly supervised learning. The first row are the coarse and fine grey value labels automatically generated by the VRBC combined with different binary vessel mask segmentations, i.e. from left to right being original segmentation method in the VRBC, SVS-net with training data generated by the original segmentation method, SVS-net with training data by manual annotation; the second row of results are test cases of the corresponding networks trained with different grey value labels.}
\end{figure}

\section{Conclusion and Discussion}
To efficiently remove background artefacts and mixed Gaussian-Poisson noises for XCA vessel extraction, we propose a detail-preserving RPCA-UNet with a patch-wise spatiotemporal SR via sparse feature selection, which can not only achieve uninformative feature pruning and Gaussian-Poisson denoising but also selectively enhance vessel features from the backgrounds. The experimental results show superior performance in both vessel extraction and vessel segmentation in an accurate and efficient way.

To the best of our knowledge, RPCA-UNet is the first neural network to implement an automatic weakly supervised vessel recovery from dynamic and complex backgrounds in XCA. Specifically, the heterogeneous grey value vessel layers automatically produced by the VRBC method\cite{qin2019accurate} are used as the training data. Such grey value labels contain the main branches of vessels, enabling RPCA-UNet to learn the greyscale and motion information of the whole vessel network. After that, RPCA-UNet can combine the information provided by the grey value labels and the characteristics of RPCA-UNet to achieve a great effect of vessel extraction. Moreover, we compared different training strategies with fine grey value labels where almost all the distal branches are annotated, and coarse grey value labels where only the major vessels and relatively thick vessel branches are annotated. The comparison results show that RPCA-UNet trained by coarse labels perform better than that trained by fine labels. Specifically, the RPCA-UNet trained by fine labels introduces significantly more noises, which is assumed to result from the overfitting in the trained neural network. Therefore, the proposed weakly supervised learning can not only largely reduce the labour and time spent on labelling data, but also improve the generalization ability of RPCA-UNet. 

To achieve a better detail-preserving vessel extraction, future research can explore more effective pooling layers\cite{singh2020eds} and interpretable\cite{huang2021interpretable} CLSTM network in the patch-wise spatiotemporal SR module for selecting sparse feature to improve the restoration of heterogeneous vessel profiles. For distal vessel detection, applying a self-attention mechanism to improve the inter-class discrimination and intra-class aggregation abilities\cite{mou2021cs2} can help unrolling network in accurately classifying the vessel pixels in the easily confused regions between the distal branches and the background.

\section*{Acknowledgements}
The authors thank all the cited authors for providing the source codes used in this work and the anonymous reviewers for their valuable comments on the manuscript.


%

\ifCLASSOPTIONcaptionsoff
  \newpage
\fi



%

%

\bibliographystyle{IEEEtr}
\bibliography{bare_jrnl}
%





\end{document}